\documentclass[ pra,letter,  paper,superscriptaddress]{revtex4-2}
\usepackage{graphicx,amsmath,amssymb,amsfonts,latexsym,xcolor,dcolumn,bm}
\usepackage{appendix}
\usepackage[colorlinks]{hyperref}
\usepackage{color}

\begin{document}

\global\long\def\eqn#1{\begin{align}#1\end{align}}
\global\long\def\vec#1{\overrightarrow{#1}}
\global\long\def\ket#1{\left|#1\right\rangle }
\global\long\def\bra#1{\left\langle #1\right|}
\global\long\def\bkt#1{\left(#1\right)}
\global\long\def\sbkt#1{\left[#1\right]}
\global\long\def\cbkt#1{\left\{#1\right\}}
\global\long\def\abs#1{\left\vert#1\right\vert}
\global\long\def\cev#1{\overleftarrow{#1}}
\global\long\def\der#1#2{\frac{{d}#1}{{d}#2}}
\global\long\def\pard#1#2{\frac{{\partial}#1}{{\partial}#2}}
\global\long\def\re{\mathrm{Re}}
\global\long\def\im{\mathrm{Im}}
\global\long\def\dd{\mathrm{d}}
\global\long\def\ddd{\mathcal{D}}
\global\long\def\avg#1{\left\langle #1 \right\rangle}
\global\long\def\mr#1{\mathrm{#1}}
\global\long\def\mb#1{{\mathbf #1}}
\global\long\def\mc#1{\mathcal{#1}}
\global\long\def\tr{\mathrm{Tr}}
\global\long\def\dbar#1{\stackrel{\leftrightarrow}{\mathbf{#1}}}

\global\long\def\nth{$n^{\mathrm{th}}$\,}
\global\long\def\mth{$m^{\mathrm{th}}$\,}
\global\long\def\non{\nonumber}

\newcommand{\bU}{{\bf{U}}}
\newcommand{\bV}{{\bf{V}}}
\newcommand{\bW}{{\bf{W}}}
\newcommand{\bd}{{\bf d}}
\newcommand{\hr}{\hat{\br}}
\newcommand{\bM}{\bf{M}}
\newcommand{\bv}{{\bf v}}
\newcommand{\hbp}{\hat{\bp}}
\newcommand{\hq}{\hat{q}}
\newcommand{\hp}{\hat{p}}
\newcommand{\ha}{\hat{a}}
\newcommand{\had}{{a}^{\dag}}
\newcommand{\ad}{a^{\dag}}
\newcommand{\hsig}{{\hat{\sigma}}}
\newcommand{\nt}{\tilde{n}}
\newcommand{\itf}{\sl}
\newcommand{\eps}{\epsilon}
\newcommand{\bsig}{\pmb{$\sigma$}}
\newcommand{\beps}{\pmb{$\eps$}}
\newcommand{\bmu}{\pmb{$ u$}}
\newcommand{\balpha}{\pmb{$\alpha$}}
\newcommand{\bbeta}{\pmb{$\beta$}}
\newcommand{\bgamma}{\pmb{$\gamma$}}
\newcommand{\bu}{{\bf u}}
\newcommand{\bpi}{\pmb{$\pi$}}
\newcommand{\bSig}{\pmb{$\Sigma$}}
\newcommand{\be}{\begin{equation}}
\newcommand{\ee}{\end{equation}}
\newcommand{\bea}{\begin{eqnarray}}
\newcommand{\eea}{\end{eqnarray}}
\newcommand{\sss}{_{{\bf k}\lambda}}
\newcommand{\ssss}{_{{\bf k}\lambda,s}}
\newcommand{\dip}{\langle\sigma(t)\rangle}
\newcommand{\dipp}{\langle\sigma^{\dag}(t)\rangle}
\newcommand{\sig}{{{\sigma}}}
\newcommand{\sigd}{{\sigma}^{\dag}}
\newcommand{\sigz}{{\sigma_z}}
\newcommand{\ra}{\rangle}
\newcommand{\la}{\langle}
\newcommand{\om}{\omega}
\newcommand{\Om}{\Omega}
\newcommand{\pa}{\partial}
\newcommand{\bR}{{\bf R}}
\newcommand{\bx}{{\bf x}}
\newcommand{\br}{{\bf r}}
\newcommand{\bE}{{\bf E}}
\newcommand{\bH}{{\bf H}}
\newcommand{\bB}{{\bf B}}
\newcommand{\bP}{{\bf P}}
\newcommand{\bD}{{\bf D}}
\newcommand{\bA}{{\bf A}}
\newcommand{\bek}{{\bf e}\rmk}
\newcommand{\rmk}{_{{\bf k}\lambda}}
\newcommand{\rk}{_{{\bf k}_1{\lambda_1}}}
\newcommand{\rkk}{_{{\bf k}_2{\lambda_2}}}
\newcommand{\rkz}{_{{\bf k}_1{\lambda_1}z}}
\newcommand{\rkkz}{_{{\bf k}_2{\lambda_2}z}}
\newcommand{\bsij}{{\bf s}_{ij}}
\newcommand{\bk}{{\bf k}}
\newcommand{\bp}{{\bf p}}
\newcommand{\epso}{{1\over 4\pi\eps_0}}
\newcommand{\bS}{{\bf S}}
\newcommand{\bL}{{\bf L}}
\newcommand{\bJ}{{\bf J}}
\newcommand{\bI}{{\bf I}}
\newcommand{\bF}{{\bf F}}
\newcommand{\bsub}{\begin{subequations}}
\newcommand{\esub}{\end{subequations}}
\newcommand{\baline}{\begin{eqalignno}}
\newcommand{\ealine}{\end{eqalignno}}
\newcommand{\Ep}{{\bf E}^{(+)}}
\newcommand{\Em}{{\bf E}^{(-)}}
\newcommand{\hbx}{{\hat{\bf x}}}
\newcommand{\hby}{{\hat{\bf y}}}
\newcommand{\hbz}{{\hat{\bf z}}}
\newcommand{\bep}{\hat{\bf e}_+}
\newcommand{\bem}{\hat{\bf e}_-}
\newcommand{\orange}[1]{{\color{orange} {#1}}}
\newcommand{\cyan}[1]{{\color{cyan} {#1}}}
\newcommand{\blue}[1]{{\color{blue} {#1}}}
\newcommand{\yellow}[1]{{\color{yellow} {#1}}}
\newcommand{\green}[1]{{\color{green} {#1}}}
\newcommand{\red}[1]{{\color{red} {#1}}}
\title{Photon Angular Momentum and Zero-Point Oscillations}
\author{Peter W. Milonni}
\affiliation{Department of Physics and Astronomy, University of Rochester, Rochester, New York 14627 USA}
\author{G. Jordan Maclay}
\affiliation{Quantum Fields LLC, St Charles, Illinois 60174 USA}
\begin{abstract}
Radiation from a localized, oscillating charge distribution can have angular momentum that cannot be explained in classical electrodynamics. We consider the simplest example---electric dipole radiation of a single 
photon---and show that this angular momentum  is attributable to zero-point oscillations in unexcited states of the dipole source. 
\end{abstract}

\maketitle
\section{Introduction}
Some features of the angular momentum of electromagnetic radiation seem paradoxical \cite{dicke,baranabov,barnettpar}.
For example, Dicke \cite{dicke} considered it ``paradoxical that a field oscillating in a mode which on grounds of symmetry contains no angular momentum should carry angular momentum when quantized," and showed that ``this angular momentum [results] from zero-point oscillations in other modes." He considered it ``quite meaningless, in the case of a quantized field, to speak of the angular momentum of a mode of the field. One must always speak of the angular momentum of the total field which depends in a complex way on the oscillations of all the modes of the field.'' This difference between the classical and quantum theories has been discussed most notably in papers by Heitler \cite{heitler} and DeWitt and Jensen \cite{dewitt} and in well-known textbooks \cite{heitlerbook,blatt,jackson}.

Analyses related to this ``paradoxical" feature of photon angular momentum \cite{dicke,heitler,dewitt,heitlerbook,blatt,jackson} have employed the general theory of field angular momentum formulated in terms of vector spherical harmonics. In this paper we consider, without recourse to vector spherical harmonics or other formal aspects of the general theory of multipole radiation, the angular momentum of a photon emitted in an electric dipole transition. This is the simplest case for which Dicke's ``paradox" can be addressed and shown to be resolved by the effect of zero-point oscillations. The zero-point oscillations in our analysis are associated with unexcited states of the dipole {\sl source}, whereas Dicke \cite{dicke} focused on the zero-point oscillations and energy of unoccupied states of the quantized {\sl field}. 

We begin in the following section by deriving a slightly more general form of an equation obtained by Landau and Lifshitz \cite{landau} for the angular momentum of an electric dipole field. This derivation avoids use of the troublesome Abraham radiation reaction force assumed by Landau and Lifshitz and does not require a time average. In Section \ref{sec:classical} we consider briefly the classical theory of the angular momentum of an electric dipole field based on this formula. A quantum-mechanical approach to the angular momentum of a single photon emitted in an electric dipole transition is presented in Section \ref{sec:quantum}. We show how a field ``which on grounds of symmetry contains no angular momentum" in classical electrodynamics can ``carry angular momentum" \cite{dicke} as a consequence of quantum-mechanical zero-point oscillations. We conclude with brief remarks in Section \ref{sec:discuss}. 

\section{Angular momentum of an electric dipole field}\label{sec:angmom}
We begin by briefly reviewing a derivation by Landau and Lifshitz \cite{landau}. Consider the angular momentum $\br\times\bp$ of a particle with radial position vector $\br$ and linear momentum $\bp$. The rate of change of angular momentum ${\bf M}_{\bd}$ for an electric dipole moment $\bd=e\br$ acted upon only by its radiation reaction force $\bF=(2e/3c^3)\stackrel{...}{\bd}$ is
\be
\frac{d\bM_d}{dt}=\br\times\frac{d\bp}{dt}=\br\times\bF=\frac{2}{3c^3}e\br\times\stackrel{...}{\bd}=
\frac{2}{3c^3}\bd\times\stackrel{...}{\bd}=\frac{2}{3c^3}\sbkt{\frac{d}{dt}(\bd\times\ddot{\bd})-\dot{\bd}\times\ddot{\bd}}.
\ee
For motion stationary in time, the time average
\be
\overline{\frac{d\bM_d}{dt}}=-\frac{2}{3c^3}\overline{\dot{\bd}\times\ddot{\bd}},
\ee
where the overline denotes the time average. This is the equation obtained by Landau and Lifshitz for the time-averaged rate at which the particle loses angular momentum due to radiation. By conservation of angular momentum \cite{baranabov}, the time-averaged rate of change of angular momentum $\bM$ of the field is
\be
\overline{\frac{d\bM}{dt}}=-\overline{\frac{d\bM_d}{dt}}
=\frac{2}{3c^3}\overline{\dot{\bd}\times\ddot{\bd}}.
\label{eq6}
\ee

We can derive an expression for the rate of change of $\bM$ in a more direct way as follows. The field angular momentum density is \cite{jackson}
\be
{\bf m}=\frac{1}{4\pi c}\br\times(\bE\times\bB),
\ee
where the electric and magnetic fields in the case of an electric dipole moment $\bd(t)$ are respectively \cite{jackson}
\be
\bE(\br,t)=\frac{1}{c^2r}\big[(\ddot{\bd}\cdot\hr)\hr-\ddot{\bd}\big]+\frac{1}{r^3}\big[3(\bd\cdot\hr)\hr-\bd]
+\frac{1}{cr^2}\big[3(\dot{\bd}\cdot\hr)\hr-\dot{\bd}\big],
\label{eq1}
\ee
\be
\bB(\br,t)=\frac{1}{cr^2}\big[\dot{\bd}\times\hr]+\frac{1}{c^2r}\big[\ddot{\bd}\times\hr\big].
\label{eq3}
\ee
$\hr=\br/r$ is the unit vector pointing from the dipole to the point $\br$ at a distance $r$ from the dipole and $\bd$ in these familiar expressions is evaluated at the retarded time $t-r/c$. The angular momentum flux (angular momentum per unit area per unit time) at point $\br$ from the dipole is $c{\bf m}$, and the rate of change of angular momentum in a differential surface element $r^2 d\Om$ is $c{\bf m}r^2d\Om$, where $\Om$ denotes solid angle. The rate of change of the angular momentum of the field from the dipole is therefore
\be
\frac{d\bM}{dt}=\frac{r^2}{4\pi}\int d\Om\big[\br\times(\bE\times\bB)\big]=
\frac{r^2}{4\pi}\int d\Om\big[\bE(\bB\cdot\br)-\bB(\bE\cdot\br)\big]
=-\frac{r^2}{4\pi}\int d\Om\bB(\bE\cdot\br),
\label{eq2}
\ee
since $\bB\cdot\br=0$. We next observe that the part 
of $\bE$ in (\ref{eq1}) that varies as $1/r$ makes no contribution to $\bE\cdot\br$. The only contribution to (\ref{eq2}) in the limit of very large $r$ (the radiation zone) therefore comes from the third term in (\ref{eq1}) and the second term in (\ref{eq3}). In the radiation zone, therefore,
\be
\frac{d\bM}{dt}=-\frac{1}{2\pi c^3r^2}\int d\Om(\ddot{\bd}\times\br)(\dot{\bd}\cdot\br).
\label{eq5}
\ee
Integration over all solid angles gives (Appendix A)
\be
\frac{d\bM}{dt}=\frac{2}{3c^3}{\dot{\bd}\times\ddot{\bd}}.
\label{eq4}
\ee
This differs from (\ref{eq6}) in that it does not involve a time average. We also note that $\dot{\bd}$ and $\ddot{\bd}$ in (\ref{eq4}) are functions of the retarded time $t-r/c$, whereas in (\ref{eq6}) they are functions of $t$.

\section{Angular momentum of radiation from an electric dipole oscillator: Classical theory} \label{sec:classical}
Consider as an example a rotating electric dipole moment
\be
\bd(t)=d_{0}[\hbx\cos\om t\pm\hby\sin\om t],
\ee
where $\hbx$ and $\hby$ are orthogonal unit vectors ($\hbx\times\hby=\hbz$) and $d_{0}$ is the amplitude of the rotating dipole moment. From Eq. (\ref{eq4}),
\be
\frac{d\bM}{dt}=\pm\hbz\frac{2\om^3}{3c^3}d_{0}^2=\pm\hbz\frac{P}{\om},
\label{eq101}
\ee
where, according to the Larmor formula,  $P=2\om^4d_0^2/3c^3$ is the radiated power. If the dipole is set into oscillation at time $t=0$, the field angular momentum after a time $t$ much greater than the radiative lifetime is
\be
{\bM}=\pm\hbz\frac{1}{\om}\int_0^tdt^{\prime} P(t^{\prime})=\pm\hbz\frac{U}{\om}=M_z\hbz,
\label{eq420}
\ee
where $U$ is the field energy. Thus,
\be
M_z=\pm\frac{U}{\om} \ \  ,  \ \ {M_z^2}=\frac{U^2}{\om^2},
\label{eq102}
\ee
a result we return to in the following section.

These results apply for circularly polarized radiation. If the dipole  simply oscillates along a single axis without rotation, it is obvious ``on grounds of symmetry" \cite{dicke} in the classical theory that the emitted field has no angular momentum.

\section{Angular momentum of photons from an electric dipole transition} \label{sec:quantum}
For a quantum-mechanical treatment of the electric dipole oscillator we will work in the Heisenberg picture and express the dipole moment as 
\be
\bd(t)=e[\hbx x(t)+\hby y(t)+\hbz z(t)],
\label{eq305}
\ee
where $e$ is the electron charge and now $\bd(t)$, $x(t)$, $y(t)$, and $z(t)$ are hermitian operators. Likewise the rate of change of angular momentum is now a hermitian, Heisenberg-picture operator defined, following (\ref{eq4}), by
\be
\frac{d\bM}{dt}=\frac{2e^2}{3c^3}\Big\{\hbx\big[\dot{y}(t_r)\ddot{z}(t_r)-\dot{z}(t_r)\ddot{y}(t_r)\big]+\hby\big[\dot{z}(t_r)\ddot{x}(t_r)-\dot{x}(t_r)\ddot{z}(t_r)\big]
+\hbz\big[\dot{x}(t_r)\ddot{y}(t_r)-\dot{y}(t_r)\ddot{x}(t_r)\big]\Big\}, \ \ \ t_r=t-r/c.
\label{eq306}
\ee

The operators $x(t),y(t)$ and $z(t)$ for oscillation at frequency $\om$ can be expressed in terms of bosonic lowering and raising operators, $a_x$ and $\ad_x$, etc., as \cite{qmbook}
\bea
x(t)&=&\sqrt{\frac{\hbar}{2m\om}}\big[a_xe^{-i\om t}+\ad_x e^{i\om t}\big],\nonumber\\
y(t)&=&\sqrt{\frac{\hbar}{2m\om}}\big[a_ye^{-i\om t}+\ad_y e^{i\om t}\big],\nonumber\\
z(t)&=&\sqrt{\frac{\hbar}{2m\om}}\big[a_ze^{-i\om t}+\ad_z e^{i\om t}\big],
\label{eq303}
\eea
where $[a_x,\ad_x]=[a_y,\ad_y]=[a_z,\ad_z]=1$ and, since the different Cartesian components of the dipole correspond to independent degrees of freedom, the lowering and raising operators for different components commute, i.e., $[a_x,a_y]=[a_x,\ad_y]=0$, etc.

It will be convenient to introduce the unit vectors
\be
{\bf e}_{\pm}=\frac{1}{\sqrt{2}}(\hbx\pm i\hby)
\ee
and the operators
\be
a_{\pm}=\frac{1}{\sqrt{2}}(a_x\mp ia_y) \ \ {\rm and} \ \  \ad_{\pm}=\frac{1}{\sqrt{2}}(\ad_x\pm i\ad_y),
\ee
such that $[a_{\pm},\ad_{\pm}]=1$, $[a_+,a_-]=[a_+,a_-^{\dag}]=[a_{\pm},a_z]=0.$ We can then express (\ref{eq305}) as 
\be
\bd(t)=\sqrt{\frac{\hbar e^2}{2m\om}}\big[a_{+}\bep +a_{-}\bem+a_z\hbz]e^{-i\om t}+{\rm h.c.},
\label{eq601}
\ee
where h.c. denotes the hermitian conjugate. The oscillator Hamiltonian can likewise be expressed as
\be
H=\hbar\om[\ad_xa_x+\ad_ya_y+\ad_za_z+3/2]=\hbar\om[\ad_{+}a_{+}+\ad_{-}a_{-}+\ad_za_z+3/2].
\ee 

The stationary states of the three-dimensional harmonic oscillator have energies $\hbar\om(n+3/2)$, $n=0,1,2,...$, orbital angular momentum quantum numbers $\ell=0,2, ..., n$ for $n$ even and $\ell=1,3, ..., n$ for $n$ odd, and $2\ell+1$ magnetic numbers $m$, $|m|=0,1, ..., \ell$ \cite{llqm}. We will consider only stationary states with $n=1$, in which case an electric dipole transition from an energy level with $\ell=1,m=0,\pm 1$ to the level with $\ell=0,m=0$ results in the emission of a single photon of energy $\hbar\om$. Note that, under a rotation by an infinitesimal angle $\delta\phi$ about the $z$ axis, $\bep$ changes by
\be
\delta\bep=\frac{1}{\sqrt{2}}(\delta\phi\hby-i\delta\phi\hbx)=-i\delta\phi\bep.
\ee
So with $z$ the quantization axis, $\bep$ is the unit vector corresponding to $m=1$ ($\delta[e^{-im\phi}]=-im\delta\phi e^{-im\phi}$). Likewise $\bem$ and $\hbz$ are associated with $m=-1$ and $m=0$, respectively. 
$\ad_{+},\ad_{-}$ and $\ad_z$ are respectively the raising operators generating oscillator states with $m=+1,m=-1$, and $m=0$. For example, if $|\psi\ra$ is the state in which all of the modes are in their ground states, 
$\ad_{+}|\psi\ra$ is the state in which there is one excitation in the 
state with $\ell=1,m=+1$ and the states with $\ell=1,m=0$ and $\ell=1,m=-1$ are unoccupied.

Using (\ref{eq305}) and (\ref{eq303}), we can write (\ref{eq306}) in terms of the raising and lowering operators: 
\be
\frac{d\bM}{dt}=\frac{2e^2\hbar\om^2}{3mc^3}\Big[\big(\ad_{+}a_{+}-\ad_{-}a_{-}\big)\hbz+\big(\ad_{-}a_z-a_{+}\ad_z\big)\bep+\big(a_{-}\ad_z-\ad_{+}a_z\big)\bem\Big].
\label{eq401}
\ee
For a stationary state $|\psi\ra$ at time $t=0$, the expectation value of $d{\bM}/dt$ at time $t\ge 0$ follows from (\ref{eq401}):
\be
\avg{\frac{{d\bM}}{dt}}=\la\psi|\frac{d\bM}{dt}|\psi\ra=\frac{2e^2\hbar\om^2}{3mc^3}\Big\la\big(\ad_{+}a_{+}-\ad_{-}a_{-}\big)\hbz+\big(\ad_{-}a_z-a_{+}\ad_z\big)\bep+\big(a_{-}\ad_z-\ad_{+}a_z\big)\bem\Big\ra,
\label{eq411}
\ee
each Heisenberg-picture operator in this equation referring to the time $t$.

If, for instance, $|\psi\ra$ is the state for which no mode of the oscillator is excited, $\la d{\bM}/dt\ra=0$. This follows simply from the fact that $a_{+}|\psi\ra=a_{-}|\psi\ra=a_z|\psi\ra=0$. In other words, there is no zero-point average angular momentum \cite{dewitt}.

Suppose the state at time $t=0$ is $|\psi_{+}\ra$, such that 
$\ad_{+}a_{+}|\psi_{+}\ra=|\psi_{+}\ra$ and $a_{-}|\psi_{+}\ra=a_z|\psi_{+}\ra=0$. Then only the $m=+1$ state is occupied and
\be
\avg{\frac{{d\bM}}{dt}}=\la\psi_{+}|\frac{d\bM}{dt}|\psi_{+}\ra=\frac{2e^2\hbar\om^2}{3mc^3}\la\psi_{+}|\ad_{+}(t)a_{+}(t)|\psi_{+}\ra\hbz=\frac{2e^2\hbar\om^2}{3mc^3}\la\ad_{+}(t)a_{+}(t)\ra\hbz.
\label{eq430}
\ee
The mean oscillator energy decreases exponentially at the radiative rate $R=2e^2\om^2/3mc^3$ \cite{jackson2}:
\be
\hbar\om\la\ad_{+}(t)a_{+}(t)\ra=\hbar\om\la\ad_{+}(0)a_{+}(0)\ra e^{-Rt}=\hbar\om e^{-Rt},
\ee
i.e.,
\be
\la\ad_{+}(t)a_{+}(t)\ra=e^{-Rt}.
\ee
Therefore, from (\ref{eq430}),
\be
\la{\bM}(t)\ra=\hbar\hbz(1-e^{-Rt}).
\label{eq520}
\ee
For times $t$ much greater than the radiative lifetime $1/R$,
\be
\la{\bM}(t\gg 1/R)\ra=\hbar\hbz=\frac{\la U\ra}{\om}\hbz,
\ee
where $\la U\ra=\hbar\om$ is the expectation value of the emitted photon energy. If the initial state is instead the singly excited state $|\psi_{-}\ra$ such that $\ad_{-}a_{-}|\psi_{-}\ra=|\psi_{-}\ra$ while $a_{+}|\psi_{-}\ra=a_z|\psi_{-}\ra=0$, we obtain in similar fashion, for this initial $m=-1$ state,
\be
\la{\bM}\ra=-\hbar\hbz=-\frac{\la U\ra}{\om}\hbz
\ee
for $t\gg 1/R$. These results are consistent with the classical expressions (\ref{eq420}) and (\ref{eq102}) for circular polarization. 

For the $m=0$ state $|\psi_0\ra$ at $t=0$, in contrast, $\la d{\bM}/dt\ra=0$. This follows from (\ref{eq411}), since $a_{\pm}|\psi_0\ra=\la\psi_0|\ad_{\pm}=0$. This is just what is expected in classical theory ``on grounds of symmetry" \cite{dicke}: a dipole oscillating along a single line has no angular momentum and can give no angular momentum to the field. 

We conclude, therefore, that for an initial 
oscillator state $\ell=1,m$ of energy $\hbar\om$, the mean angular momentum $\la\bM\ra$ ending up in the field satisfies 
\be
\frac{\la\bM\ra}{\la U\ra}=\frac{m}{\om} \ \ , \ \ 
\frac{\la\bM\ra^2}{\la U\ra^2}=\frac{m^2}{\om^2}.
\label{eq440}
\ee

\section{Effect of zero-point oscillations}\label{sec:zpe}
As noted, (\ref{eq440}) is consistent with expectations from classical theory. One might furthermore expect, for a stationary state 
$\ell=1,m$ at time $t=0$, that at times $t\gg 1/R$,
\be
\frac{\la{\bM}^2\ra}{\la U^2\ra}=\frac{m^2}{\om^2}.
\label{eq441}
\ee
But this expectation would be incorrect, since the square of the angular momentum in quantum theory is $\ell(\ell+1)\hbar^2=2\hbar^2$, i.e.,
\be
\frac{\la\bM^2\ra}{\la U^2\ra}=\frac{\ell(\ell+1)\hbar^2}{\hbar^2\om^2}=\frac{2}{\om^2},
\label{eq442}
\ee
which is greater than \eqref{eq441} for all allowed values $(0,\pm 1)$ of $m$. Jackson attributes the difference between \eqref{eq441} and \eqref{eq442} to the uncertainty principle: ``If the $z$ component of angular momentum of a single photon is known precisely, the uncertainty principle requires that the other components be uncertain, with mean square values such that [\eqref{eq442}] holds" \cite{jackson}. Griffiths and Schroeter, similarly, observe that the magnitude $\sqrt{\ell(\ell+1)}\hbar$ of the angular momentum exceeds (for $\ell>0$)  the maximum $z$ component ($\ell\hbar$): ``Evidently you can’t get the angular momentum to point perfectly along the z direction ... to do that you would have to know all three components simultaneously, and the uncertainty principle ... says that’s impossible" \cite{qmbook2}. 

So the uncertainty principle explains why (\ref{eq441}) is incorrect. We can go further and obtain explicitly the correct result (\ref{eq442}) in a way that does not invoke the uncertainty principle but rather follows Dicke's perspective \cite{dicke}. For this purpose we return to (\ref{eq401}) and write it more compactly as
\be
\frac{d{\bM}}{dt}=R\hbar{\bf F}(t),
\ee
where the Heisenberg-picture operator
\be
{\bf F}(t)=\big[\ad_{+}(t)a_{+}(t)-\ad_{-}(t)a_{-}(t)\big]\hbz+\big[\ad_{-}(t)a_z(t)-a_{+}(t)\ad_z(t)\big]\bep+\big[a_{-}(t)\ad_z(t)-\ad_{+}(t)a_z(t)\big]\bem.
\ee
If the oscillator is excited only after a time $t=0$, the mean-square field  angular momentum for $t\ge 0$ is
\be
\la{\bM}^2\ra=R^2\hbar^2\int_0^tdt^{\prime}\int_0^tdt^{\prime\prime}\big\la{\bf F}(t^{\prime})\cdot{\bf F}(t^{\prime\prime})\big\ra.
\label{eq530}
\ee
For $t^{\prime},t^{\prime\prime}\ge 0$,
\be
\big\la{\bf F}(t^{\prime})\cdot{\bf F}(t^{\prime\prime})\big\ra=\la{\bf F}^2(0)\ra e^{-R(t^{\prime}+t^{\prime\prime})}
\ee
as a consequence of the radiative decay of the dipole moment at the rate $R$. For $t\gg 1/R$ we obtain, from (\ref{eq530}),
\bea
\la{\bM}^2\ra&=&\hbar^2\la{\bf F}^2(0)\ra=
\hbar^2\big\la\ad_{+}a_{+}\ad_{+}a_{+}
-2\ad_{+}a_{+}\ad_{-}a_{-}+\ad_{-}a_{-}\ad_{-}a_{-}+\ad_{-}a_{-}a_z\ad_z-2\ad_{-}\ad_{+}a_za_z-2a_{+}a_{-}
\ad_z\ad_z\nonumber\\
&&\mbox{} + a_{+}\ad_{+}\ad_{z}a_{z}
+\ad_{z}a_{z}a_{-}\ad_{-}+\ad_{+}a_{+}a_{z}\ad_{z}\big\ra,
\eea
where all the raising and lowering operators refer to time $t=0$.

Suppose, for example, that the oscillator is initially in 
the  $\ell=1,m=0$ stationary state $|\psi_0\ra$. Then $a_z|\psi_0\ra=|\psi_0\ra$, $a_{\pm}|\psi_0\ra=0$, and
\be
\la{\bM}^2\ra=\la\psi_0|{\bM}^2|\psi_0\ra=\hbar^2\big\la a_{+}\ad_{+}\ad_{z}a_z+a_{-}\ad_{-}\ad_za_z\big\ra=2\hbar^2\la\ad_{z}a_z\ra=2\hbar^2,
\ee
\be
\frac{\la{\bM}^2\ra}{\la U^2\ra}=\frac{2\hbar^2}{(\hbar\om)^2}=\frac{\ell(\ell+1)}{\om^2},
\label{eq501}
\ee
since $\la a_{+}\ad_{+}\ad_za_z\ra=\la a_{-}\ad_{-}\ad_za_z\ra=\la\ad_za_z\ra=1$. This result for the state $|\psi_0\ra$ is in sharp contrast to the prediction of classical theory for a state in which the dipole moment undergoes linear oscillation along the $z$ axis. Classically, there is no angular momentum in the radiation from the dipole. Quantum mechanically, the mean angular momentum of the single-photon field vanishes, but not its mean square. This mean-square angular momentum is attributable to the zero-point oscillations in the two {\sl unexcited} circular polarization modes, i.e., to the fact that $\la a_{+}\ad_{+}\ra=\la a_{-}\ad_{-}\ra=1$. To quote Dicke again, the total angular momentum of the field ``depends in a complex way on the oscillations of all the modes of the field" \cite{dicke}, {\sl including modes which, though unexcited, have zero-point energy and undergo zero-point oscillations.}

The same result (\ref{eq501}) applies as well to the states $|\psi_{\pm}\ra$. For the state $|\psi_{+}\ra$, for example, we obtain, from the fact that $a_z|\psi_{+}\ra=a_{-}|\psi_{+}\ra=0$,
\bea
\la{\bM}^2\ra=\hbar^2\la\ad_{+}a_{+}\ad_{+}a_{+}+\ad_{+}a_{+}a_z\ad_z\ra=\hbar^2\big(\la\ad_{+}a_{+}\ad_{+}a_{+}\ra+\la\ad_{+}a_{+}\ra\la a_z\ad_z\ra\big)=\hbar^2\big(1+\la a_z\ad_z\ra\big)
=2\hbar^2.
\label{eq505}
\eea
In this case the zero-point oscillations of the mode linearly polarized along $z$ give a contribution $\la a_z\ad_z\ra=1$, without which we would obtain only half the result (\ref{eq505}).

We have, then, for any stationary state $\ell=1,m$ of the oscillator at $t=0$, the relation
\be
\frac{\la\bM^2\ra}{\la U^2\ra}=\frac{2}{\om^2}=\frac{\ell(\ell+1)}{\om^2}.
\ee
This is a special case, for $N=1$ (one photon) and $\ell=1$ (electric dipole transition), of the general multipolar result quoted by DeWitt and Jensen \cite{dewitt} and Jackson \cite{jackson}:
\be
\frac{\la\bM^2\ra}{\la U^2\ra}=\frac{N^2m^2+N\ell(\ell+1)-m^2}{N^2\om^2}.
\ee
The classical result (\ref{eq441}) for $m=\pm 1$ follows when $N\gg 1$ \cite{jackson}. For $m=0$, $\la{\bM}^2\ra/\la U\ra\rightarrow 0$ as $N\rightarrow\infty$, consistent with the classical prediction that a linearly polarized field carries no angular momentum.

Our analysis has employed the dipole oscillator model, a simple and often useful model for an electric dipole transition. It can be brought into more quantitative agreement with the more generally applicable two-level-atom model for an electric dipole transition by replacing the ratio $e^2/m$ by $fe^2/m$, where \cite{cray}
\be
f=2m\om|d_0|^2/e^2\hbar
\ee
is the oscillator strength for the transition with frequency $\om$ and transition dipole matrix element $d_0$. This implies the replacement
\be
\frac{2e^2\hbar\om^2}{3mc^3}\rightarrow \frac{4|d_0|^2\om^3}{3\hbar c^3}\hbar=A\hbar,
\label{eq410}
\ee
where $A$ is the Einstein $A$ coefficient for the rate of spontaneous emission for a transition of frequency $\om$ and transition dipole matrix element $d_0$ \cite{acoeff}. Then $R$ in (\ref{eq520}), for instance, is replaced by $A$. With this replacement our results in Section \ref{sec:quantum} are in agreement with the more detailed analysis by Franke and Barnett \cite{barnett2} of the dynamics of mean angular momentum transfer from the atom to the field in spontaneous emission. We can likewise replace $R$ everywhere in the present section by $A$ without affecting at all our conclusions about the role of zero-point oscillations. 

\section{Remarks}\label{sec:discuss}
The fact that zero-point oscillations of unexcited modes can result in a nonvanishing mean-square angular momentum has a clear physical explanation. The angular momentum in the $z$ direction, for instance, depends on products of $x$ and $y$ displacements, as in (\ref{eq306}). Classically, if there is a nonvanishing component along $y$ but not along $x$, the angular momentum about $z$ vanishes. Quantum mechanically, however, zero-point oscillations along the $x$ direction result in a nonvanishing mean-square angular momentum about $z$, since these zero-point oscillations have nonvanishing mean squares. 

This is made especially clear in an example simpler than that considered by Dicke or in this paper. Consider a plane monochromatic wave of frequency $\om$ propagating in a direction $\hat{\bk}$. Let ${\bf e}_{{\bk}1}$ and ${\bf e}_{{\bk}2}$ be two real, orthogonal unit vectors corresponding to two orthogonal linear polarizations, such that ${\bf e}_{{\bk}1}\times{\bf e}_{{\bk}2}=\pm\hat{\bk}$. The angular momentum about the $\hat{\bk}$ direction may be shown to be \cite{mandelwolf}
\be
{\bM}=i\hbar\hat{\bk}\big(\ad_2a_1-\ad_1a_2\big),
\ee
where $a_j$ ($\ad_j$), $j=1,2$, is the photon annihilation (creation) operator for the mode linearly polarizaed along ${\bf e}_{{\bf k}j}$. If, for instance, the mode with linear polarization along ${\bf e}_{{\bf k}1}$ is in a state in which there is exactly one photon, and the mode with linear polarization along ${\bf e}_{{\bf k}2}$ is in its vacuum state $|0\ra$ of no photons, then $\la{\bM}\ra=0$, since $a_2|0\ra=\la 0|\ad_2=0$. However,
\be
\la{\bM}^2\ra=\hbar^2\big\la \ad_2a_2a_1\ad_1+\ad_1a_1a_2\ad_2-\ad_2\ad_2a_1a_1-\ad_1\ad_1a_2a_2\big\ra=\hbar^2\big\la\ad_1a_1a_2\ad_2\big\ra=\hbar^2\big\la\ad_1a_1\big\ra=\hbar^2,
\ee
since $a_2\ad_2|0\ra=|0\ra$, a fact attributable to the zero-point oscillations and energy $\frac{1}{2}\hbar\om$ of mode 2.
Of course this is as expected from a different perspective: the linearly polarized mode 1 can be considered to be an equal superposition of left- and right-circularly polarized modes, each contributing $\frac{1}{2}\hbar^2$ to $\la{\bM}^2\ra$.

\section*{Acknowledgment}
We thank Paul R. Berman, Ralf Menzel, and Kanu Sinha for helpful remarks.
\\ \\
The authors have no conflicts to disclose.

\appendix
\section{Derivation of Equation (\ref{eq4})}
The $x$ Cartesian component of the integral (\ref{eq5}) over all solid angles has the form, for $\bU=\bU(t-r/c)$ and $\bV=\bV(t-r/c)$,
\bea
(\bW)_x&=&\int d\Om(\br\times \bU)_x(\bV\cdot\br)
=\int_0^{\pi}d\theta\sin\theta\int_0^{2\pi}d\phi\big(yU_z-zU_y)(\bV\cdot\br)\nonumber\\
&=&r^2\int_0^{\pi}d\theta\sin\theta\int_0^{2\pi}d\phi\big(U_z\sin\theta\sin\phi-U_y\cos\theta\big]\big)\big(
V_x\sin\theta\cos\phi+V_y\sin\theta\sin\phi+V_z\cos\theta\big)\nonumber\\
&=&r^2\int_0^{\pi}d\theta\sin\theta\int_0^{2\pi}\big(U_zV_y\sin^2\theta\sin^2\phi-U_yV_z\cos^2\theta\big)
=-\frac{4\pi}{3}r^2(U_yV_z-V_yU_z)\nonumber\\
&=&\frac{4\pi}{3}r^2(\bV\times\bU)_x,
\eea
and likewise for the $y$ and $z$ components of $\bW$, i.e., $\bW=(4\pi/3)r^2\bV\times\bU$. Equation (\ref{eq4}) follows when we put $\bU=\ddot{\bd}(t-r/c)$ and $\bV=\dot{\bd}(t-r/c)$.

\end{document}